\documentclass[a4paper,twoside]{article}%
\newread\tmp

\usepackage{todonotes}
\usepackage{xcolor}
\usepackage{listings}
\usepackage{booktabs}
\usepackage{rotating}
\usepackage{enumitem}

\usepackage{epsfig}
\usepackage{subcaption}
\usepackage{calc}
\usepackage{amssymb}
\usepackage{amstext}
\usepackage{amsmath}
\usepackage{amsthm}
\usepackage{multicol}
\usepackage{pslatex}
\usepackage{apalike}
\usepackage{algorithm2e}
\usepackage[bottom]{footmisc}
\usepackage{SCITEPRESS}     

\begin{document}

\title{Qualitative In-Depth Analysis of GDPR Data Subject Access Requests and Responses from Major Online Services}

\author{\authorname{Daniela Pöhn\sup{1}\orcidAuthor{0000-0002-6373-3637} and Nils Gruschka\sup{2}\orcidAuthor{0000-0001-7360-8314}}
\affiliation{\sup{1}University of the Bundeswehr Munich, Research Institute CODE, Munich, Germany}
\affiliation{\sup{2}Department of Informatics, University of Oslo, Oslo, Norway}
\email{daniela.poehn@unibw.de, nilsgrus@ifi.uio.no}
}

\abstract{The European General Data Protection Regulation (GDPR) grants European users the right to access their data processed and stored by organizations. Although the GDPR contains requirements for data processing organizations (e.\,g., understandable data provided within a month), it leaves much flexibility. In-depth research on how online services handle data subject access request is sparse. Specifically, it is unclear whether online services comply with the individual GDPR requirements, if the privacy policies and the data subject access responses are coherent, and how the responses change over time. 
To answer these questions, we perform a qualitative structured review of the processes and data exports of significant online services to (1) analyze the data received in 2023 in detail, (2) compare the data exports with the privacy policies, and (3) compare the data exports from November 2018 and November 2023. The study concludes that the quality of data subject access responses varies among the analyzed services, and none fulfills all requirements completely.}

\onecolumn \maketitle \normalsize \setcounter{footnote}{0} \vfill

\section{\uppercase{Introduction}}

The modern world has become data-centric and dominated by large enterprises that provide digital services and collect large amounts of personal data. Following this, data has become a commodity exploited and traded for commercial advantage, giving behavioral insights for advertisements. 
To re-balance the power over personal data, the European Union's (EU) General Data Protection Regulation (GDPR)~\cite{gdpr} came into force on May 25, 2018. Among others, the GDPR grants individuals (i.\,e., data subjects) rights to access their data and explain its usage in an electronic and understandable form. Since then, similar regulations, such as the California Consumer Privacy Act, have been introduced worldwide. 
Research has mainly concentrated on other GDPR-related topics, such as cookie banners and their effects on ad networks. Research on data subject access request (DSAReq; short: request) and data subject access response (DSARes; short: response) is still rare. 
However, research concerning data subject access with actual account data may provide insights into their compliance with the GDPR, among other things. Hence, a qualitative in-depth analysis of leading online services is essential.

Therefore, we initiated requests at Amazon, Google, Facebook, Microsoft, LinkedIn, Apple, and WhatsApp as the leading online services and retrieved and analyzed the responses. 
This paper makes the following contributions: a qualitative analysis of the data request process of leading online services in detail; a matching of the responses with the corresponding privacy policies; a comparison of the responses from the beginning of the GDPR and today. 
With these contributions, we are addressing the following research questions:
    Do the selected online services comply with the GDPR concerning DSAReq and DSARes? To what extent do the privacy policies match the data responses? How do the responses from November 2018 and 
    2023 differ for the selected online services?

We summarize the background in Section~\ref{sec:background} and contrast the related work with our approach. In Section~\ref{sec:methodology}, we outline our qualitative method. Section~\ref{sec:analysis} comprises the study of the DSAReq workflows, the DSARes, a comparison of 2018 and 2023, and an evaluation of the privacy policies. Based on the results, we discuss our findings (see Section~\ref{sec:discussion}) and conclude the paper.

\section{\uppercase{Related Work}}
\label{sec:background}


The GDPR is the primary legal framework for data protection within the European Union, including Norway, Iceland, and Liechtenstein. 
The regulation defines, among other things, fundamental principles and definitions, obligations for organizations that process personal data (``data controllers''), and rights of individuals whose data is processed (``data subjects'').

The main research topics in the area of GDPR are tracking, cookies, and consent, 
dark patterns
, and compliance and enforcement. 
Fewer approaches address the topic of DSAReq and DSARes. \cite{10.1145/3491102.3501947} conducted a user study with ten participants, in which each participant filed four to five data access requests. The authors notice non-compliance and low-quality responses. Similarly, \cite{9229806} present a user study with 13 households who request DSARes from their loyalty program providers. The authors conclude that the responses should deliver detailed information to prevent mistrust. 
DSARes donations by users can be applied in research \cite{
laura_boeschoten_2021_4472606,10.5117/CCR2022.2.002.BOES,Boeschoten2021,10.1080/19312458.2022.2109608}. However, the data has to be cleaned from private data. \cite{10.1007/978-3-031-68024-3_7} introduce a method to create synthetic DSAR datasets, whereas \cite{Peters2023} outline the different variants of DSAReqs at Instagram. In contrast, \cite{10.1145/3600160.3605064} analyze DSARes concerning conformity, finding differences in the type of service and request method. \cite{10.1007/978-3-031-34444-2_9} automate performing DSAReqs.

Based on the related work, only a few authors focus on data subject access requests and responses. 
A qualitative in-depth analysis of actively used accounts related to Art. 5 GDPR is missing. To notice changes in the DSARes between years and users, comparing the responses from several years and users could help shed light on the practices of online services. 

\section{\uppercase{Method}}
\label{sec:methodology}




We select the big five tech companies for our analysis: Google, Amazon, Apple, Meta (i.\,e., Facebook and WhatsApp), and Microsoft. They are leading in e-commerce, consumer electronics, online advertising, online searches, and social media. Consequently, we assume that they potentially collect a large amount of data. We also select \cite{linkedin} as a business-focused social media platform with over 850 million registered members from over 200 countries.
We utilize two actively used accounts for each service. The accounts were created and used primarily in Europe. User $A$ (female) is privacy-concerned with non-standard operating systems, script blockers, and similar, but shares some work-related data and regularly buys items at Amazon. User $B$ (male) uses standard devices and no additional measures. They have used their accounts for several years, as shown in Table~\ref{tab:account}. We expect realistic results using these inartificial and long-lived accounts and can analyze minimization, retention time, and other requirements. 
Both users consented to the analysis. As the DSARes contain personal identifiable information, the users analyzed their own data. This procedure is in line with the ethical boards of the universities.
\begin{table}[!htpb]
\centering
\caption{Account creation year for both main users.}
\label{tab:account}
\begin{tabular}{lcc}
\toprule
\textbf{Online service} & \textbf{User A} & \textbf{User B} \\ \midrule
Amazon   & 2006 & 2008 \\
Apple    & 2007 & 2015 \\
Facebook & 2007 & 2009 \\
Google   & 2009 & 2012 \\
LinkedIn & 2012 & 2009 \\
WhatsApp & 2013 & 2015 \\  
\bottomrule       
\end{tabular}
\end{table}
\label{sec:documentation}
To document the DSAReqs, we establish a template 
based on the results published by \cite{10.1145/3600160.3605064,10.1007/978-3-031-68024-3_7}. 
\label{sec:evaluation-criteria}
Next, the data is evaluated. We chose a manual step-by-step process since automatic tools do not provide the required detail. 

\section{\uppercase{Results}}
\label{sec:analysis}

In the following, we present the results per online service. 
The summarized results are presented in Tables~\ref{tab:template1} and~\ref{tab:template2} in the Appendix. A comparison between these online services is being made in Section~\ref{sec:discussion}.

\subsection{Amazon}


The user receives a DSARes with at least 47 folders, including those for account settings, advertisements, Alexa, app store, Audible, devices, digital content, Prime Video, Kindle, notifications, payments, and retail. One DSARes had 212 folders and 15,906 files. 
Receiving an overview of the data might be difficult without a primary HyperText Markup Language (HTML) page.
We found old email addresses in several files that the users had changed. At several locations, such as \texttt{Amazon-Music} or retail (i.\,e., their web store), searches requests (incl. search terms and timestamps) are stored for the whole lifetime of the account. The amount of data generally is high, including old data from 2012 to 2015. However, the data is, at least in parts, incomplete. 
In the second step, we highlight specific files and folders. The \texttt{Alexa} folder includes audio files and transcriptions, though the user did not use Alexa but an early version of a FireTV. The \texttt{Smart Home} folder contains Alexa voice-enabled devices, including all smartphones. \texttt{Devices.Registration} contains all devices -- multiple times, partly with the wrong status. 
\texttt{Digital.Content.Whispersync} lists all actions and states while reading, such as marked text, started and stopped reading, reading speed, and comments. \texttt{Digital.PrimeVideo.LocationData} consists of various locations the users have been to. One user with infrequent app usage could find around 20 entries per day. This implies that Amazon apps may collect data during non-usage phases.
\texttt{OutboundNotification.AmazonApplicationUp dateHistory} shows all updates of Amazon apps since 2020, with active debugging status. 
\texttt{Retail.AuthenticationTokens} includes authentication records with several old sessions that seem to be still active. The shopping profile is somewhat amusing for both users (notice that User B is male), as the shopping profile contains `female' and `shoes'.
Amazon's privacy policy provides detailed information about the data it collects, processes, stores, and uses without informing users about the duration or retention. The policy is coherent with the in-depth analysis. However, the content is relatively condensed (approximately 4,000 words in English at the time of the study), so it may not be easily understandable for non-technical users. 
Comparing the folders and files from 2018 and 2023 reveals many changes. Several files were introduced in 2023, such as \texttt{Advertising}, \texttt{Audible}, and \texttt{AccountSettings.PrivacyPreferences.Con sents}. These can partly be derived from changes in the offered services. However, several files are newly added that contain data from 2018 and before, such as \texttt{Digital.Borrows.2} (data from 2014), \texttt{Digital-Ordering.2} (data from 2012), and \texttt{PrimeVideo.Viewing-History} (data from 2014). In total, we find 44 such new files with old content.

\subsection{Apple}


The DSAReq from Apple has 3 to 11 ZIP files. These include about 40 to 75 files in 20 to 40 directories of up to 5 levels nested. 
The file formats are well-suited to automatic processing. However, the readability for humans, especially non-technical persons, is extremely poor.
The received data suggests that Apple stores only necessary data and retains it for a reasonable duration. 
Changes to the account information (\texttt{Apple ID Account Information}) or AppStore transactions (\texttt{Store Transaction History}) seem to be kept infinitely, but this is probably appropriate.
Remarkably, certain information, like email addresses, phone numbers, and credit card numbers, are shown partly redacted, like \texttt{j***d**@gmail.com}. 
Apple's privacy policy is approximately 4,000 words long and contains detailed information about the collected data, e.\,g., account information, devices, and usage information, which seems to follow our data analysis. 
The policy also mentions that Apple receives data from and shares data with other parties, but it is phrased very vaguely (``Apple may ...''). Our DSARes did not include any information regarding shared data. The retention time is not noted, which makes a comparison difficult.
The main structure of the responses has not changed from 2018 to 2023. However, new data has been included in the 2023 version. For example, recovery devices and devices with Apple messaging are added to the folder related to other data. 
The AppleID account and device information folder includes the latest files \texttt{Apple ID Device Information.csv} and \texttt{Data \& Privacy Request History.csv} containing data from 2018. Similarly, the folders \texttt{Game Center} and \texttt{Information about Apple Media Services} are newly introduced, containing data from, for example, 2011. 

\subsection{Facebook}


If the HTML format is chosen during the request, the main HTML page 
is 
similar to Facebook. 
The number of folders depends on conversions and media uploads, with a minimum of 58 folders for information about ads, apps and websites, connections, files, logged information, personal information, preferences, security and login information, and activities. 
Information about ads comprises data on advertisers based on activities or information, though partly not fitting to the user, a deleted blog page, and connected websites and apps that were removed in 2018. The timing (GDPR coming into effect) is also notable, as the users did not use these online services and apps then. The logged information contains the location with postal code and timezone, though not intentionally added, and interactions starting in 2013, among others. 
We again recognize the location in security and login information, however, less accurate than in the location data. Logins, sessions, types of sessions, terminated sessions, geolocation, browser fingerprints, known devices, and browser cookies since 2012 or 2011 are logged. 
Meta's privacy policy concerns Facebook, as well as several other Meta services that may have additional privacy policies. The overall structure is user-friendly, explaining every item, even with videos (around 13,000 words in English). 
We notice that Facebook/Meta claims to log much data, including the name of the network carrier, language, timezone, mobile number, IP address, download speed, network capacity, information about nearby devices, WiFi hotspots, and mouse movements. This could partly not be verified with our responses. 
Information about cookies is rather generic, and the usage of shadow profiles is hinted at. However, we could not find information related to the retention time.
Many files and folders have been renamed, while new files, for example, \texttt{supervision}, \texttt{files}, \texttt{preferences} (*), \texttt{logged\_information} (*),  and \texttt{your-problem-reports}, have been included. Both (*) files seem interesting, as these probably were stored already in 2018. 

\subsection{Google}


The DSARes of Google (excluding the content of Google Drive and the uploaded YouTube videos) contains approx.\ 100 files in approx.\ 75 directories. 
For browsing the DSARes, an HTML page is included that groups the files into approx. 50 categories. All files contain a short explanation of their content. 
Like Apple, the overall impression is that the amount of data and storage time are appropriate for most categories. For example, the recent logins (which also contain IP addresses and user agent information) are only stored for approx.\ half a year while the history of installation and purchases from the Google Play Store are stored infinitely. Also, activities like search history or a list of watched YouTube videos are stored infinitely. This behavior can be configured in the account dashboards to no storage of activities or deletion after 18 months for instance. 
Google's privacy policy is similar in length to Meta's (approximately 14,000 words). It is nicely presented with illustrations, videos, and links to the aforementioned dashboards for configuring information access and deletion. 
However, regarding retention duration, the policy remains rather vague.
Comparing the responses from 2018 and 2023, we notice that several folders and files have been renamed. 

\subsection{LinkedIn}


The DSARes from LinkedIn consist only of a set of CSV files. No human-readable data format or navigation help is provided. The number of files depends upon the features used (e.\,g., job search), but it is much smaller than the other services discussed above. 
After analyzing the content, we found the following noticeable aspects: Per device/browser, only the last login is stored by LinkedIn and presumably only for two years. The file \texttt{Ads clicked} contains a long (up to 300 entries in our cases) list of timestamps (from the last two years) and ``ad ids''. 
The DSARes contains ``facts'' that the users have not explicitly provided but have been inferred by LinkedIn, e.\,g., gender or date of birth. 
LinkedIn's privacy policy is approximately 6,000 words long, a mean size compared to the other services analyzed. The policy is nicely written, with explanations and links to further information. Information on the storage of logins complies with the data found in the DSARes. 
The policy does not mention the two years observed for login information.
The comparison of the responses does not show many changes. 

\subsection{Microsoft}
\label{sec:microsoft}


We observed severe issues with requesting the data exports. This included finding the request form, different paths to requests depending on the account type (private or business), and authentication codes sent via email that never arrived.
The one received DSARes consists of the file \texttt{ProductAndServiceUsage.csv} with date, end-date, aggregation, app name, and app publisher. Further data can be requested separately for Skype, OneDrive, Microsoft 365, and Microsoft Teams. 
Data about the account, usage, and additional services, such as email, is not included. Therefore, we conclude that Microsoft is not compliant concerning the request's possibility and completeness.
Microsoft's privacy policy is the longest, with around 44,000 words. This can partly be explained as it includes the policies of various products. According to the privacy policy, Microsoft stores data about interactions, such as device and usage data, interests, content consumption data, searches and commands, voice data, texts, images, contacts and relationships, social data, location data, and other input. However, we could not find that in the responses. The cookie information seems incomplete, as the third-party cookie information contains only two generic sentences. Furthermore, we noticed broken links in the policy. Finally, information related to retention is missing.
In 2018, the only data received was a short extract about the Skype service. 
No data about the account or activities was included. However, the data in 2023 is not much more. 
Although OneDrive is not used, it appears twice. However, no information about emails or other similar information can be found.

\subsection{WhatsApp}


The DSARes from WhatsApp contains only six HTML files (plus an \texttt{index.html} file), making it very easy to browse the information. 
The included data is limited to a minimum. 
The privacy policy for Whats\-App is rather lengthy (approx. 16,000 words). It lists detailed data types that are collected and stored. This includes data types not being part of our DSARes, e.\,g., battery level and signal strength. 
The comparison between both responses reveals only a few changes. The most significant difference is that the data is now shown in a more user-friendly way by using HTML. Also, more data on the account registration is provided.

\section{\uppercase{Evaluation}}
\label{sec:discussion}


Based on the results, presented in Section~\ref{sec:analysis} and summarized in Tables~\ref{tab:template1} and~\ref{tab:template2} in the Appendix, 
we compare the online services. Table~\ref{tab:comp} shows an overview of the results.

\emph{DSAReq and DSARes:} DSA requests and responses were possible at most services, although utilizing desktop browsers was mandatory within LinkedIn. We observed several issues at Microsoft that led to only one DSARes being available.

\emph{Completeness:} Completeness is never given, as we cannot proof that the online service has provided us with all the data. We are only sure that Microsoft did not provide all the data as, for example, the registration data is missing.

\emph{Correctness:} Although we evaluate correctness, we did not use controlled data \cite{10.1145/3600160.3605064,10.1007/978-3-031-68024-3_7}, but historical data to receive realistic results. Thereby, we cannot strictly compare input and output. However, we found suspicious data at Amazon (outdated addresses deleted previously) and Facebook (data about a page that seems to be active, although deleted previously).

\emph{Understandable:} Concerning understandable data, we rate JSON as machine-readable and HTML as understandable. 
WhatsApp, LinkedIn, and Facebook fulfill this criterion, while Amazon, Apple, and Microsoft are considered incomprehensible. 

\emph{Data minimization:} For data minimization, we rate the historical data found during the analysis that is detailed in Section~\ref{sec:history}. 
We noticed that none of the online services fulfill all the criteria. Microsoft performs worst (only one fulfilled), while WhatsApp performs best (four fulfilled). 

\begin{table}[!htpb]
\renewcommand{\arraystretch}{1.2}
\centering
\caption{Comparison of the online services based on the evaluation criteria.}
\vspace{0.2em}
\label{tab:comp}
\begin{tabular}{lccccc}
\toprule
& \textbf{DSAR} & \textbf{Com.} & \textbf{Cor.} & \textbf{Und.} & \textbf{Min.} \\ 
\midrule
Am.    &  +  &  $\circ$  &  $\circ$ & -- &  -- \\
Ap.     &  +  & $\circ$  &  + &  -- &  $\circ$  \\
FB  &  +  & $\circ$  &  $\circ$ &  +  &  -- \\
Go.   &  +  &  $\circ$  &  + &  $\circ$  &  $\circ$  \\
LI &  $\circ$  &  $\circ$  &  + &  +  &  $\circ$  \\
Mi. &  -- &  -- &  + &  -- &  $\circ$  \\
WA  &  +  &  $\circ$  &  + &  +  &  +  \\ \bottomrule
\end{tabular}
\par\medskip
\scriptsize	
DSAR = DSAReq and DSARes, Com. = Completeness,\\ Cor. = Correctness, Und. = Understandable, Min. = Data minimization. \\
Am. = Amazon, Ap. = Apple, FB = Facebook, Go = Google,\\ LI = LinkedIn, Mi. = Microsoft, WA = WhatsApp. \\
$+$ = fulfilled, $\circ$ = partly fulfilled or unknown, $-$ = not fulfilled.
\end{table}
\label{sec:history}

\begin{figure}
\centering
\includegraphics[width=\linewidth]{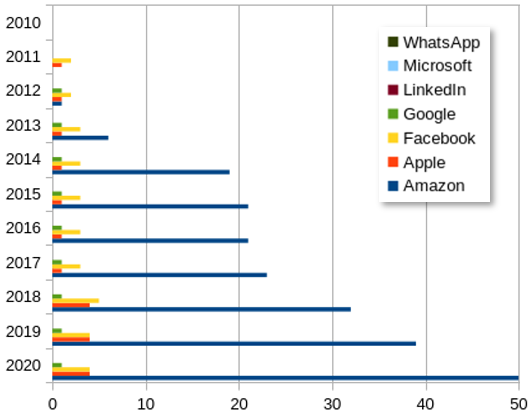}
\caption{Data with history data by year and online service.}
\label{fig:data-year}
\end{figure}

The amount of historical data indicates if a service complies with the data minimization and storage limitation principles (see Art. 5 GDPR). 
The results of the data until 2020 (note that newer data is not seen as historical) can be seen in Figure~\ref{fig:data-year}. We have found no historical data from 2010 or before, though some services were used at that time (see Table~\ref{tab:account}). However, data, like cookies and searches, is included from 2011 and 2012. 
Generally, most online services presented us with more data in 2023 due to newly launched services. Based on the talk by \cite{datenmuell}, we were not surprised to find more data in the DSARes of Amazon that should have been in the one from 2018. We had the same issue with Apple. For Facebook, we assume the same.


For the most part, the analyzed privacy policies are easy to understand and sometimes even include explanatory videos (see Google and Facebook). However, they do not provide the details to compare them stepwise with the DSARes, such as the exact data type. Microsoft is the only online service that clearly does not give enough data based on its privacy policy.
The retention time (see Arts. 13 and 15) is typically not given in the privacy policies. As outlined by \cite{10.1007/978-3-030-33752-0_6}, the GDPR is vague in its interpretation of deletions (for example, concerning timeliness and deletion method). We notice the results in the DSARes, as visualized in Figure~\ref{fig:data-year} (see Section~\ref{sec:history}). 




\section{\uppercase{Conclusion}}
\label{sec:conclusion}

The EU's GDPR grants individuals rights to access their data and have its usage explained in an electronic and understandable form. This paper provides the first qualitative in-depth analysis of the requests and responses from major online services by analyzing their current data subject access requests and responses, comparing 2018 and 2023, and comparing their privacy policies. 
Overall, the data subject access process is satisfactory for nearly all services, but the amount of data varies greatly between the different services. 
Also, regarding the accessibility and understandability of responses, we experienced large differences between the services. 
Further, comparing the responses from 2018 and 2023 revealed that Amazon and Apple did not provide all the data in their earlier responses. Finally, vague information made mapping the responses with the privacy policies impossible. 



\bibliographystyle{apalike}
{\small
\bibliography{GDPR}}

\section*{\uppercase{Appendix}}

\label{sec:appendix-template}










Table~\ref{tab:template1} describes the path to the request and the request itself, while Table~\ref{tab:template2} contains the notification, download, and data (without date and time).

\begin{sidewaystable*}[hp]
\centering
\caption{Template 1/2 containing data about the path to the request and the request itself.}
\label{tab:template1}
\begin{tabular}{p{3cm}p{2cm}p{2cm}p{2.8cm}p{2.5cm}p{2.2cm}p{2cm}p{2cm}}
\toprule
\textbf{Categories}  & \textbf{Amazon} & \textbf{Apple}   & \textbf{Facebook}  & \textbf{Google}  & \textbf{LinkedIn}  & \textbf{Microsoft}  & \textbf{WhatsApp} \\ \midrule
\textit{(a) Path to request}  &  &  &  &  &  &  &  \\
Starting point  & Web + App & Web + App  & Web + App  & Web + App & Web + App  & Web + App  & App \\
Steps to request   & 5  & 6  & 7--10  & 7  & 5  & n & 4   \\
Further help required  & No  & No   & No  & No  & Partly  & Yes  & No  \\
Observations  & User-friendly & Easier in browser & User-friendly  & Easier in browser  & Not possible within the app & Not intuitive  & User-friendly \\ \midrule
\textit{(b) Request} &  &  &  &  &  &  &  \\
Form of request & Form  & Form & Form & Form & Button & Multiple  & Button  \\
Authentication for request (*)  & L + 2FA & L + 2FA  & L  & L + Pwd   & L + MFA  & L + MFA  & L  \\
Selection options  & Categories & Categories  & Categories and notification & Categories, file type, frequency \& destination  & none  & services  & none \\
Observations  & User-friendly  & User-friendly     & User-friendly  & User-friendly  & User-friendly  & Partly not possible & User-friendly     \\ \bottomrule
\end{tabular}
\par\medskip
\scriptsize (*)	
\textbf{L:} Logged into account; \textbf{2FA:} Second-factor authentication; \textbf{MFA:} Multi-factor authentication (at least 3 authentication factors);\newline \textbf{Pwd:} Additional password.
\end{sidewaystable*}

\begin{sidewaystable*}[hp]
\centering
\caption{Template 2/2 containing data about the notification and data.}
\label{tab:template2}
\begin{tabular}{p{3cm}p{2cm}p{2cm}p{2cm}p{3.5cm}p{2cm}p{2cm}p{2cm}}
\toprule
\textbf{Categories}  & \textbf{Amazon} & \textbf{Apple}   & \textbf{Facebook}  & \textbf{Google}  & \textbf{LinkedIn}  & \textbf{Microsoft}  & \textbf{WhatsApp} \\ \midrule
\multicolumn{2}{l}{\textit{(c) Notification and download}}  &  &  &  &  &  &  \\
Form of information about data  & Email  & Email  & Email/In-App  & Email  & Email  & Download  & In-App  \\
Time between request and data available & \textless 3 days  & \textless 1 week  & \textless 3 days  & \textless 1 day  & \textless 2 days  & \textless 5 min & \textless 3 days  \\
Steps to data  & 4  & 5  & 3  & 4 & 4  & 1  & 2  \\
Authentication for data (*)  & L + 2FA & L + 2FA  & L  & L + Pwd  & L + MFA & L  & L \\
Observations   & User-friendly   & User-friendly     & Email might be sent  & User-friendly  & User-friendly & Not intuitive  & User-friendly \\ \midrule
\textit{(d) Data}   &  &  &  &  &  &  &  \\
Data formats  & CSV, EML, JPEG, JSON, PDF, TXT, WAV, README & CSV, ICS, JSON    & JSON or HTML, TXT, JPG, PNG, GIF  & HTML, CSV, JSON, TXT,  PDF, MBOX, VCF, ICS, README, JPG, PNG, atom, DIC, DOCX, FRC, GIF, ICO, MP4, PPTX, XLSX, XML & CSV  & CSV  & HTML, JSON, PNG   \\
Data type  & Machine-readable  & Machine-readable & Both  & Both  & Machine-readable  & None  & Both  \\
Folders/Categories  & 212  & 24  & 296  & 133  & 1  & 0  & 14   \\
Number of files  & 15906  & 51  & 616  & 5414  & 33  & 1  & 5  \\
Observations   &  &  &  &  &  & not complete  &  \\ \bottomrule
\end{tabular}
\par\medskip
\scriptsize (*)	
\textbf{L:} Logged into account; \textbf{2FA:} Second-factor authentication; \textbf{MFA:} Multi-factor authentication (at least 3 authentication factors);\newline \textbf{Pwd:} Additional password.
\end{sidewaystable*}

\end{document}